\documentstyle[referee]{mn}
\include{epsf}
\newif\ifAMStwofonts

 \def\etal{et al.}
 \def\ie{{i.e.,}\,}
 \def\eg{{e.g.,}\,}

 \def\la{\hbox{\raise.5ex\hbox{$<$}
     \kern-1.1em\lower.5ex\hbox{$\sim$}}}
 \def\ga{\hbox{\raise.5ex\hbox{$>$}
     \kern-1.1em\lower.5ex\hbox{$\sim$}}}
 \def\sun{\odot}
 \def\ms{M_{\sun}}
 \newcommand\eqref[1]{(\ref{#1})}

\ifoldfss
  \ifCUPmtlplainloaded \else
    \NewTextAlphabet{textbfit} {cmbxti10} {}
    \NewTextAlphabet{textbfss} {cmssbx10} {}
    \NewMathAlphabet{mathbfit} {cmbxti10} {} 
    \NewMathAlphabet{mathbfss} {cmssbx10} {} 
  \fi
  \ifAMStwofonts
    \ifCUPmtlplainloaded \else
      \NewSymbolFont{upmath} {eurm10}
      \NewSymbolFont{AMSa} {msam10}
      \NewMathSymbol{\upi}     {0}{upmath}{19}
      \NewMathSymbol{\umu}     {0}{upmath}{16}
      \NewMathSymbol{\upartial}{0}{upmath}{40}
      \NewMathSymbol{\leqslant}{3}{AMSa}{36}
      \NewMathSymbol{\geqslant}{3}{AMSa}{3E}

    \fi
  \fi
\fi 

\ifnfssone
  \newmathalphabet{\mathit}
  \addtoversion{normal}{\mathit}{cmr}{m}{it}
  \addtoversion{bold}{\mathit}{cmr}{bx}{it}
  \newmathalphabet{\mathbfit} 
  \addtoversion{normal}{\mathbfit}{cmr}{bx}{it}
  \addtoversion{bold}{\mathbfit}{cmr}{bx}{it}
  \newmathalphabet{\mathbfss} 
  \addtoversion{normal}{\mathbfss}{cmss}{bx}{n}
  \addtoversion{bold}{\mathbfss}{cmss}{bx}{n}
  \ifAMStwofonts
    \ifCUPmtlplainloaded \else
      \UseAMStwoboldmath
      \makeatletter
      \new@mathgroup\upmath@group
      \define@mathgroup\mv@normal\upmath@group{eur}{m}{n}
      \define@mathgroup\mv@bold\upmath@group{eur}{b}{n}
      \edef\UPM{\hexnumber\upmath@group}
      \new@mathgroup\amsa@group
      \define@mathgroup\mv@normal\amsa@group{msa}{m}{n}
      \define@mathgroup\mv@bold\amsa@group{msa}{m}{n}
      \edef\AMSa{\hexnumber\amsa@group}
      \makeatother
      \mathchardef\upi="0\UPM19
      \mathchardef\umu="0\UPM16
      \mathchardef\upartial="0\UPM40
      \mathchardef\leqslant="3\AMSa36
      \mathchardef\geqslant="3\AMSa3E
    \fi
  \fi
\fi 

\ifnfsstwo
  \DeclareMathAlphabet{\mathbfit}{OT1}{cmr}{bx}{it}
  \SetMathAlphabet\mathbfit{bold}{OT1}{cmr}{bx}{it}
  \DeclareMathAlphabet{\mathbfss}{OT1}{cmss}{bx}{n}
  \SetMathAlphabet\mathbfss{bold}{OT1}{cmss}{bx}{n}
  \ifAMStwofonts
    \ifCUPmtlplainloaded \else
      \DeclareSymbolFont{UPM}{U}{eur}{m}{n}
      \SetSymbolFont{UPM}{bold}{U}{eur}{b}{n}
      \DeclareSymbolFont{AMSa}{U}{msa}{m}{n}
      \DeclareMathSymbol{\upi}{0}{UPM}{"19}
      \DeclareMathSymbol{\umu}{0}{UPM}{"16}
      \DeclareMathSymbol{\upartial}{0}{UPM}{"40}
      \DeclareMathSymbol{\leqslant}{3}{AMSa}{"36}
      \DeclareMathSymbol{\geqslant}{3}{AMSa}{"3E}
    \fi
  \fi
\fi 

\ifCUPmtlplainloaded \else
  \ifAMStwofonts \else 
    \def\upi{\pi}
    \def\umu{\mu}
    \def\upartial{\partial}
  \fi
\fi

\title{Does the plasma composition affect the long term evolution of
relativistic jets?}
\author[L.~Scheck et al.]
{L.~Scheck,$^1$ M.~A.~Aloy,$^1$ J.~M.~Mart\'{\i},$^2$ J.~L.~G\'omez$^3$ and 
E.~M\"uller$^1$ \\
 $^1$Max-Planck-Institut f\"ur Astrophysik, Garching 85748, Germany \\
 $^2$Departamento de Astronom\'{\i}a y Astrof\'{\i}sica, Universidad
     de Valencia, 46100 Burjassot, Spain \\
 $^3$Instituto de Astrof\'{\i}sica de Andaluc\'{\i}a, 18080 Granada, Spain}
\date{Accepted 2001 August ??.
      Received 2001 August ??;
      in original form 2001 August ??}

\pagerange{\pageref{firstpage}--\pageref{lastpage}}
\pubyear{2001}

\begin{document}

\maketitle

\label{firstpage}

\begin{abstract}
We study the influence of the matter content of extragalactic jets on
their morphology, dynamics and emission properties. For this purpose
we consider jets of extremely different compositions including pure
leptonic and baryonic plasmas. Our work is based on two-dimensional
relativistic hydrodynamic simulations of the long-term evolution of
powerful extragalactic jets propagating into a homogeneous
environment.  The equation of state used in the simulations accounts
for an arbitrary mixture of electrons, protons and electron-positron
pairs. Using the hydrodynamic models we have also computed synthetic
radio maps and the thermal Bremsstrahlung X-ray emission from their
cavities.

 Although there is a difference of about three orders of magnitude in
the temperatures of the cavities inflated by the simulated jets, we
find that both the morphology and the dynamic behaviour are almost
independent on the assumed composition of the jets.  Their evolution
proceeds in two distinct epochs.  During the first one
multidimensional effects are unimportant and the jets propagate
ballistically.  The second epoch starts when the first larger vortices
are produced near the jet head causing the beam cross section to
increase and the jet to decelerate.  The evolution of the cocoon and
cavity is in agreement with a simple theoretical model.  The beam
velocities are relativistic ($\Gamma \simeq 4$) at kiloparsec scales
supporting the idea that the X-ray emission of several extragalactic
jets may be due to relativistically boosted CMB photons.  The radio
emission of all models is dominated by the contribution of the hot
spots.  All models exhibit a depression in the X-rays surface
brightness of the cavity interior in agreement with recent
observations.

\end{abstract}

\begin{keywords}
hydrodynamics -- relativity -- plasmas -- ISM: jets and outflows --
radiation mechanisms: thermal -- galaxies: jets.
\end{keywords}

\section{Introduction}

The standard model for powerful jets associated with extragalactic
radio sources \cite{BR74} assumes that the energy
radiated in the radio lobes of such sources is produced in the active
nuclei of their host galaxies, the central engine being a supermassive
black hole surrounded by an accretion disc. The accretion process
fuels a couple of twin supersonic jets which transport away bulk
kinetic energy from the neighbourhood of the central black hole to the
lobes, \ie from scales of the Schwarzschild radius of the supermassive
black hole $R_S=2GM_{\rm bh}/c^2 = 3\cdot10^{13} \cdot (M_{\rm
bh}/10^8\ms)\,$cm, to kiloparsec scales. This bulk kinetic energy is
dissipated by shocks within the beam and (mostly) at the jet terminal
shocks where electrons are accelerated and radiate via synchrotron and
inverse Compton mechanisms. Typical lifetimes and kinetic powers of
powerful radio sources are $\approx 10^7\,$y and $10^{44} -
10^{47}\,$erg s$^{-1}$, respectively \cite{RS91,Da95}.

 Among the problems that still remain open even after more than 30
years of research is the composition of extragalactic jets.  Within
the standard model \cite{BR74} jets are made of a relativistic plasma
that contains relativistic electrons and thermal protons
($ep$-plasma).  On the other side, based on equipartition arguments
Kundt \& Gopal-Krishna \shortcite{KG80} claim that extragalactic jets
consist of beams of extremely relativistic electrons and positrons
($e^\pm$-plasma) with almost no ions. Ultimately, the composition of
jets is tightly related to their formation mechanisms. As discussed by
Celotti \& Blandford \shortcite{CB01}, electromagnetically dominated
outflows, as those generated by the extraction of spin energy of the
black hole, will become pair dominated jets with a low baryonic
pollution. Jets generated from the accretion disk by hydromagnetic
winds will be made of baryonic plasma. Probably, both processes are
operating simultaneously in nature.  Sol, Pelletier \& Asseo
\shortcite{SPA89} propose a {\it two-flow} model where the jet
consists of a beam of relativistic particles (pair plasma) surrounded
by a Newtonian or mildly relativistic ($ep$-plasma) wind accelerated
from the disk. It is also possible that $ep$-jets become pair-loaded
later on, \eg by interactions with high energy photons from the disk
corona -- \eg Begelman, Blandford \& Rees \shortcite{BBR84} -- or by
proton-proton collisions at parsec scale -- Anyakoha, Okeke \& Okoye
\shortcite{AOO88} --.

 Observations show that a number of jets are highly polarised. This is
a further argument in favour of jets made of e$^\pm$ pairs as there is
no internal Faraday rotation or depolarisation (the rotation produced
by the electrons is compensated by that of the positrons), while an
$ep$-plasma gives rise to these effects. If jets were dominated by
$ep$-plasmas, observational limits on the degree of Faraday rotation
and depolarisation imply thermal electron densities $<10^{-3}\,
$cm$^{-3}$ -- \eg Walker, Benson \& Unwin \shortcite{WBU87} -- or a
minimum energy cut-off for the electrons $\approx 50\,$MeV
\cite{Wa77}.  This suggests a lack of thermal (cold) matter, at least,
in the most polarised sources.  However, as the upper bounds on the
thermal electron density derived from the depolarisation argument
depend on the magnetic field structure, they might be underestimated
if there are field reversals within the VLBI beam.  Finally, the
detection of circular polarisation at parsec scales in several radio
sources \cite{Wetal98,HW99} suggests that, in general, extragalactic
radio sources are mainly composed of $e^\pm$-plasma.  The argument is
based on the fact that circular polarisation is produced by Faraday
conversion, which requires that the energy distribution of the jet
emitting particles extends to very low energies. This in turn
indicates that $e^\pm$-pairs are an important component of the jet
plasma.

Sikora \& Madejski \shortcite{SM00} conclude that X-ray observations
of blazars associated with OVV quasars impose strong constraints on
the e$^\pm$ pair content of jets of radio-loud quasars. According to
these authors, pure electron-positron pair jets can be excluded
because they may produce too much soft X-rays, while pure
electron-proton jets may emit too few non-thermal X-ray
radiation. Therefore, only jets may be viable where the number density
of electron-positron pairs is much larger than the proton number
density, but small enough for protons still being dynamically more
important. Hirotani \etal \shortcite{Hietal00} find that parsec-scale
jets cannot be dominated by baryons, if the electron density
determined by the amount of synchrotron self-absorption necessary for
an optically thick jet component, is of the same order as the one
derived from the kinetic luminosity.

For jets composed of a pair plasma additional questions arise. How can
they maintain their stability, as for a given particle density they
are lighter and hence less stable than jets made of an
electron-proton plasma? Can they transport enough energy to fuel the
observed kiloparsec scale radio lobes?  Concerning the stability of
electron-positron plasma flows, Sol \etal \shortcite{SPA89}
demonstrate that the {\it two-flow} model is stable against
excitation of electrostatic waves as long as the magnetic field is
larger than a critical value $B_c$.  In their {\it two-flow} model,
the relativistic beam is responsible for the VLBI jet and the observed
superluminal motion, while the slower wind gives rise to the
kiloparsec scale jet.  Achatz \& Schlickeiser \shortcite{AS93} also
consider the stability of $e^\pm$-plasmas against the excitation of
electromagnetic waves when the magnetic field exceeds $B_c$. They find
that the plasma may be stable over large distances provided plasma
waves are damped thermally. The problem of the energy supply to large
scales with a limited number of particles is turned around by Celotti
\shortcite{Ce98}. She argues that the jets of low power radio sources
mainly consist of $e^\pm$-plasma, while those of powerful radio-loud
quasars are made of $ep$-plasma. However, measurements of the
circular polarisation of VLBA-jets of several FR\,II sources by
Wardle \etal \shortcite{Wetal98} favour $e^\pm$-jets also in powerful
sources (at least, at parsec scales).

 Another important and not yet solved question concerns the impact of
the composition on the morphology and dynamics of jets from sub-parsec
to kiloparsec scales. Observations indicate that jets are relativistic
at parsec scales -- \eg Laing \shortcite{La96} --, then decelerate
until they become sub-relativistic or mildly relativistic -- \eg Bridle
\etal \shortcite{Betal94} -- at kiloparsec scales with advance speeds
of the terminal hotspots in the range $0.01c$ to $0.1c$
\cite{LPR92,Da95}, where $c$ is the speed of light. At large scales
the observed morphologies and deceleration of powerful radio sources
is governed by interaction with the external medium.  Several
theoretical models have considered the gross features of kiloparsec
scale jets, \ie their long term evolution.  Begelman \& Cioffi
\shortcite{BC89} -- BC89 hereafter -- consider a simple model to
describe the evolution of the cocoon. Self-similar expansion is
suggested by Falle \shortcite{Fa91}.  Komissarov \& Falle
\shortcite{KF98} explore the large-scale flow caused by classical and
relativistic jets in a uniform external medium. They find that jets
with finite initial opening angles are recollimated by the high
pressure in the cocoon and that the flow becomes approximately
self-similar at large times.

Numerical investigations have also addressed the kiloparsec scale
regime. Two-dimensional Newtonian hydrodynamic simulations of
axisymmetric light jets were performed by Cioffi \& Blondin
\shortcite{CB92} in order to understand the evolution of the cocoon,
and were compared with the simple analytic theory of BC89. The grid
resolution is quite high (15 zones per jet radius), but the
simulations do only cover a relatively short period of the cocoon's
evolution.  The Newtonian simulations of Hooda, Mangalam \& Wiita
\shortcite{HMW94} cover the evolution of axisymmetric extragalactic
jets up to $10^8\,$y. Their simulations include isothermal atmospheres
with a density stratification given by a power-law, surrounded by an
even hotter, but less dense intra-cluster medium where the jets
accelerate and collimate.  Hooda \& Wiita \shortcite{HW96} extended
the results of Hooda \etal \shortcite{HMW94} to three-dimensions, but
their simulations cover only a distance of 35 (initial) jet radii, and
thus are too short to shed light on the long term evolution of real
sources.

Mart\'{\i}, M\"uller \& Ib\'a\~{n}ez (1998) -- MMI98 hereafter --
studied the long term evolution of powerful extragalactic jets on the
basis of relativistic hydrodynamic simulations (up to an evolutionary
time of $3 \cdot 10^6\,$y with a relatively low numerical
resolution). The results are compared and interpreted with a simple
generalisation of the model of BC89. They find an evolution divided
into two epochs. After a transient initial stage, the jet's evolution
is dominated by a strong deceleration. The jet advance speed becomes
as small as $0.05c$ due to the degradation of the beam flow by means
of internal shocks and the broadening of the beam cross section near
the hotspot.

The present work extends the investigations of MMI98 to much later
evolutionary times and also addresses the question whether the content
of thermal matter in powerful kiloparsec jets does influence their
morphology, dynamics and emission properties. Like MMI98 we assume
that the dynamics of the jets is dominated by the thermal plasma --
\eg Sikora \& Madejski \shortcite{SM00} -- and, thus, a hydrodynamic
approach is appropriate. For this purpose we have performed long-term
simulations of axisymmetric, relativistic jets of extremely different
composition, \ie jets made of a pure leptonic or baryonic plasma,
propagating into a uniform environment.

\section{Models}
\label{sec:models}

\subsection{Equation of state} 
\label{subsec:EoS}

In almost all previous jet simulations an ideal gas equation of state
(EoS) with constant adiabatic index $\gamma$ has been used.  This is a
good approximation in both the nonrelativistic ($\gamma=5/3$) and the
ultra-relativistic limit ($\gamma=4/3$).  However, when there exist
very large temperature gradients in the flow, it is more accurate to
use a EoS including a temperature dependent $\gamma$. For the
simulations that we present below temperatures range from about
$10^7\,$K in the ambient medium to $10^{13}\,$K in the hotspots.
Thus, both relativistic and nonrelativistic particles will participate
on an equal footing. Additionally, extragalactic jets are likely
composed of a mixture of particles of different masses, \ie the value
of $\gamma$ depends on the composition as protons become relativistic
at higher temperatures than electrons. 

An equation of state that describes a mixture of ideal, relativistic
Boltzmann gases has been derived by Synge \shortcite{Sy57} (see also
Komissarov \& Falle 1998, appendix A). The Synge EoS used in our
simulations includes protons, electrons and positrons. The composition
of the plasma only changes due to fluid mixing as the production or
annihilation of electron-positron pairs can be neglected due to the
low gas density in the kiloparsec scale (see, \eg Ghisellini \etal
1992). Assuming plasma neutrality, only one parameter is needed to fix
the composition, \eg the mass fraction of the leptons
$X_l=(\rho_{e^-}+\rho_{e^+})/\rho$ where $\rho_{e^-}$ and $\rho_{e^+}$
are the local rest-mass densities of electrons and positrons,
respectively. Using the Synge EoS instead of a constant-$\gamma$ EoS
requires about 50\% more computation time, because the iterative
(Newton-Raphson) computation of $T(\varepsilon,\rho,X_l)$
($\varepsilon$ being the specific internal energy) involves Bessel
functions.

\subsection{The Parameter Space}
\label{subsec:param}
The following notation holds throughout the paper.  Subscripts $b$ and
$m$ refer to the beam and the external medium, respectively. The speed
of light is set to $c=1$.

Assuming a uniform and static external medium a relativistic
jet is fully described specifying, at a given inlet of radius $R_b$,
the density contrast $\eta$ between the beam and the ambient medium,
the pressure ratio $K = P_b / P_m$ between the beam and the ambient
gas, the beam speed $v_b$ or equivalently the beam Lorentz factor $W_b
= 1/\sqrt{(1-v_b^2)}$, the beam Mach number $M_b = v_b / c_{sb}$ (where
$c_{sb}$ is the beam sound speed), and parameters that depend on the
EoS. For an ideal gas EoS only the adiabatic index $\gamma$ needs to
be specified. In this case, a relativistic jet is freely scalable in
size and density, but not in velocity due to the scale introduced by
the speed of light. Using the Synge EoS (\S\ref{subsec:EoS}) $\gamma$
is replaced by two parameters describing the chemical composition in
the jet and in the external medium, \eg the lepton mass fractions
$X_{lb}$ and $X_{lm}$, respectively.  The Synge EoS also introduces an
extra mass scale (due to the electron and proton masses), and hence an
additional (mass) parameter, \eg $m_0 = \rho_m R_b^3$.  With this set
of parameters $\{ \eta, K, M_b, W_b, m_0, X_{lb}, X_{lm} \}$ a jet
model is scalable under the transformations $t \rightarrow at, x
\rightarrow ax, \rho \rightarrow \rho/a^3$ ($a = const$). In the
following this scale freedom will not be used, but $\rho_m$ and $R_b$
will be fixed instead of $m_0$.

Current observations provide only constraints for the parameter space.
The kinetic luminosity $L_{kin}$ of a typical FR II jet is
$10^{46}{\rm erg/s}$ \cite{RS91,Da95}. The initial jet
propagation speed can be restricted by observations of CSOs (Compact
Source Object) which have linear sizes of less than 0.5\,kpc and which
most likely represent a very early evolutionary stage of a typical
powerful radio source. Observed values of the hotspot propagation
velocity are $\approx 0.2c$ \cite{OC98,Tayetal00}.

The density, temperature and composition of the external medium are
$\rho_m \approx 10^{-3} m_p/{\rm cm}^3 = 1.67 \cdot 10^{-27}{\rm
g/cm^3}$, $T_m \approx 10^7 {\rm K}$ and $X_{lm} = m_e/(m_e+m_p)
\approx 1/1837$, respectively (\eg Ferrari 1998). Values of $R_b \sim
0.5 {\rm kpc}$ can be inferred from observations of kiloparsec scale
jets. But $R_b$ is also constraint through other model parameters (see
below). There are also observational constraints on the likely value
of the beam Lorentz factor at parsec scale - $W_b \approx 10$, \eg
Ghisellini \etal \shortcite{Ghetal93} --. Furthermore, measurements of the
degree of polarisation as a function of frequency may be used to set
an upper limit on the mean thermal electron density number which is
$n_e \approx 10^{-2} \,$cm$^{-3}$ \cite{PWS79,Bu79,Ghetal92}.

We have fixed the values of $L_{kin}$, $X_{lm}$, $\rho_m$, $T_m$ and
the initial propagation speed in all our models. The latter is set
equal to
\begin{equation}
        \label{equ:v1d}
        v_j^{\rm 1d} =
        \frac{\sqrt{\eta_R}}{\sqrt{\eta_R}+1} v_b\, ,
\end{equation}
with $\eta_R = \rho_bh_b W_b^2/\rho_mh_m$.
This is the propagation velocity derived by Mart\'{\i} \etal
\shortcite{Metal97} for a pressure matched jet that moves in one
dimension only (\ie without sideways expansion). If the jet is not
pressure matched, the propagation velocity can still be computed using
the previous formula as long as the sound speed in the external medium
fulfils $c_{sm} << 1$.  The density contrast $\eta$ and the
composition of the beam ($X_{lb}$) are not directly accessible to
observations, \ie these quantities are treated as free parameters. The
beam Lorentz factor $W_b$ cannot be chosen arbitrarily, because there
are {\it forbidden} areas in the parameter space where the pressure
ratio $K$ and, most importantly, the beam temperature $T_b$ have
unrealistic values (see below). We used values of $W_b$ in the range
$6.6 - 8$.

$L_{kin}$ is obtained integrating the energy flux (see, Mart\'{\i}
\etal 1997, Eq.\,20) over the beam cross section
\begin{equation}
	\label{equ:Lkin}
	L_{kin} = (h_b W_b - 1) \rho_b W_b \pi R_b^2 v_b\, .
\end{equation}
This differs from the definition that other authors have used in the
past -- see \eg Ghisellini \shortcite{Ghi98} -- by the factor $h_b W_b -
1$ which accounts for the fact that we do not include the rest-mass
energy in the energy density, as it cannot be extracted from the beam
particles.

Having fixed $L_{kin}$, $v_j^{1d}$, $X_{lb}$ and the external medium,
the values of $\eta$, $T_b$ and $M_b$ are uniquely determined by $W_b$
and $R_b$ through relations (\ref{equ:v1d}) and (\ref{equ:Lkin}). To
clarify this point, let us define function
\begin{equation}
	C_1 := \eta h(\eta,T_b)
	     = \frac{h_m}{W_b^2}  \frac{(v_j^{1d}/v_b)^2}{(1-v_j^{1d}/v_b)^2}
\end{equation}
and
\begin{equation}
	C_2 := \eta ( h(\eta,T_b) W_b - 1 )
	     = \frac{L_{kin}}{\rho_m W_b \pi R_b^2 v_b} \, .
\label{equ:C2}
\end{equation}
Then $\eta = C_1 W_b - C_2$. For $T_b$ there exists no closed form,
instead we have the relation:
\begin{equation}
	h( C_1 W_b - C_2, T_b ) = \frac{C_1}{C_1 W_b - C_2} \, .
\end{equation}
From $T_b$ and the EoS one can compute $c_{sb}$ and thus the Mach
number $M_b$. Figure \ref{fig:pspace1} shows the dependence of $T_b$
and $\eta$ on $W_b$ and $R_b$ in the case of leptonic jets. For large
values of $W_b$ there is only a small range of values of $R_b$ around
$0.35\,$kpc where solutions exist. In fact, the asymptotic value of
the beam radius 
\begin{equation}
R_b^{\infty} = \sqrt{\frac{L_{kin}}{\rho_m\pi}} \cdot
               \frac{1-v_j^{1d}}{v_j^{1d}}
\label{eqn:Rb}
\end{equation}
is obtained when $W_b \rightarrow \infty$ and $T_b=0$.
%
%

Computing $R_b$ from (\ref{eqn:Rb}) using the independently measured
values of $L_{kin}$, $v_j^{1d}$ and $\rho_m$ from observations yields
$R_b \approx 0.35 {\rm kpc}$, which is in agreement with standard
values for kpc scale jets (\eg Ferrari 1998).  Fixing the values of
$v_j^{1d}$ and $L_{kin}$, and for a constant large $W_b$ (\ie $v_b
\approx c$), it turns out that
\begin{equation}
k := \eta h_b = const  
\end{equation}
and
\begin{equation}
(k W_b - \eta) R_b^2 = const \, .
\label{eqn:kW}
\end{equation}
The latter equation implies that for a given value of $W_b$, $\eta$
must increase with increasing $R_b$. As $k=const$ this means $h_b$ and
thus $T_b$ must decrease. Eventually, if $R_b$ increases, $T_b$
becomes negative. This explains the physically forbidden area to the
right of $R_b^{\infty}$ in Fig.~\ref{fig:pspace1}. A similar argument
holds for the forbidden area to the left: decreasing $R_b$ eventually
leads to non-physical solutions with
$\eta<0$. Figure~\ref{fig:pspace1} also shows that for a given $W_b$
there exists a maximum allowed $\eta$ and for a given $\eta$ there
exists a maximum $W_b$. This maximum $W_b$ grows increasing the value
of $v^{\rm 1d}_j$. Let us also point out that from Eq.~(\ref{equ:C2}),
increasing values of $L_{\rm kin}$ lead to larger values of $R_b$
(notice the quadratic dependence). Hence, although there exits no
solution with the seemingly reasonable values $W_b=10$, $\eta=10^{-3}$
and $R_b \approx 0.35 {\rm kpc}$ (for the chosen parameters of the
external medium, $v^{\rm 1d}_j$, and $L_{\rm kin}$), one can obtain
models with these characteristics by increasing $v^{\rm 1d}_j$ and
$L_{\rm kin}$ up to 0.24 and $1.6\cdot10^{46}{\rm erg/s}$, respectively.
Slightly smaller values of $W_b$ have been chosen to keep our
parameters the closest possible to the observed values.

Most of the previous jet simulations assumed that the jets are
pressure matched ($K=1$), because (i) there are no direct measurements
of $K$, and (ii) this reduces the parameter space. However, fixing $K$
implies that the properties of the external medium depend on the
choice of the jet parameters. Therefore, one has to adjust the
temperature of the external medium of every jet model such that
$K=1$. As we wanted to simulate typical FRII-jets in a typical
environment, the properties of the external medium are fixed and the
assumption $K=1$ is abandoned. For $K>1$ jets can still remain stable,
because they are pressure confined by the cocoon, where the pressure
is much larger than in the external medium.  In the simulated models
$K$ is in the range $1-200$. Under-pressured jets ($K<1$) cause
considerable numerical problems when they are evolved beyond $3 \cdot
10^6\,$y, because the amount of ambient gas entrained into the beam
becomes large. As the beams of our models are under-dense with respect
to the external medium, dense blobs of matter from the ambient medium
begin to pile up in front of the nozzle blocking the inflow. As we
assume axisymmetry the material piles up around the jet axis.

\subsection{Numerical models}
\label{subsec:models}

The simulations are performed with the relativistic hydrodynamic,
high-resolution shock-capturing code of MMI98.  However, instead of a
constant $gamma$-law EoS we have used the EoS described in
Sect.~\ref{subsec:EoS}. The code is well suited to solve the equations
of relativistic hydrodynamics in cylindrical coordinates, \ie for
problems with axial symmetry. It has been extensively tested and
applied previously (\eg, Mart\'{\i} \etal 1995, 1997, MMI98). Covering
a time span of up to $6.6 \cdot 10^6\,$y the set of models represent
the longest simulations of relativistic jets performed so far.

We have considered three models (Table \ref{tab:models}) including two
leptonic ($X_{lb}=1$) ones and one baryonic model ($X_{lb}=10^{-3}$;
BC).  One of the leptonic models has a factor of 100 lower density
$\eta=10^{-5}$ (LH) than the other two models with $\eta=10^{-3}$.
The corresponding number densities of thermal electrons in the beam
are $n_e \simeq 2 \cdot 10^{-5}$ and $n_e \simeq 2 \cdot 10^{-3}$,
respectively. Note that both values are below the upper limit
estimated from measurements of the degree of polarisation (see
Sect.~\ref{subsec:param}).  The low $\eta$ leptonic model has been run
in order to check how extremely low values of $\eta$ affect the long
term evolution of a relativistic jet (see, \eg Birkinshaw 1991;
Ferrari 1998).

All models have the same power and initial propagation velocity, and
they all propagate into the same homogeneous external medium with
$\rho_m=10^{-3} m_p/{\rm cm^3}$, $T_m=10^7{\rm K}$, and $X_{lm} =
m_e/(m_e+m_p) \approx 1/1837$.  As the jet power is fixed, the lower
the density the higher is the internal energy, \ie the light jet is
also the hottest ($\varepsilon \gg 1$). Note also that although the
values of the injection temperatures (Table \ref{tab:models}) seem to
be rather large, they are unanimously determined once the fiducial
conditions in the external medium, $L_{kin}$, $v_j^{1d}$ and $X_{lb}$
have been fixed.

We have not considered a light baryonic model (the counterpart of the
light leptonic model) for three reasons. First, the temperatures
within the beam and in some parts of the cocoon would reach
$10^{14}\,$K. Thus, the dynamics of the model would be very similar to
that of the hot leptonic model, \ie radiation dominated.  Second, at
temperatures of $10^{14}\,$K (\ie Lorentz factors of $\approx 10^4$)
leptons of the thermal plasma will contribute significantly to the
synchrotron radiation power.  Assuming an equipartition magnetic field
($\simeq 60 \mu$Gauss for our simulations; see
Sect.\,\ref{subsec:observations}), the synchrotron cooling time is
$t_{\rm cool}\approx 10^4 - 10^5\,$y. This time is much smaller than
the typical lifetime of the source, \ie one has to consider
synchrotron cooling in the simulations. Third,  the light baryonic
model is too hot. Its hot thermal plasma would have a detectable
emissivity which is not observed in real sources, so far (see, \eg
Celotti \etal 1998).

\begin{table}
 \caption{Parameters of the three simulated jet models.}
 \label{tab:models}
 \centering
 \begin{tabular}{@{}lccc}
 \hline
Model: & BC & LC & LH \\
 \hline
$L_{kin} \: [{\rm erg/sec}]$ 	& \multicolumn{3}{|c|}{$10^{46}$}\\
$v_j^{\rm 1d} \: [c]$		& \multicolumn{3}{|c|}{$0.2$}\\
 \hline
$X_{lb}$             & $10^{-3}$	    & $1$	& $1$\\
$\eta$		     & $10^{-3}$ 	    & $10^{-3}$ & $10^{-5}$\\
 \hline
$\varepsilon_b \: [c^2]$& $5.69 \cdot 10^{-3}$ & $0.298$   & $119$\\
$T_b \: [{\rm K}]$   & $10^{10}$            & $10^9$    & $2.37 \cdot10^{11}$\\
$M_b$		     & $16.4$	            & $2.38$    & $1.71$\\
$W_b$		     & $7.95$	            & $6.62$    & $6.62$\\
$K$		     & $1.4$		    & 91        & 217\\
$\gamma_b$	     & $1.42$	            & $1.50$    & $1.33337$\\
$R_b \: [{\rm kpc}]$ & $0.366$	            & $0.361$   & $0.342$\\
 \hline
 \end{tabular}
\end{table}


In order to track the evolution of the beam material the beam mass
fraction $X=\rho_b/\rho$ is evolved by our code by means of an
additional continuity equation. We also have to evolve the lepton
fraction $X_{l}$ because, as explained in Sect.\,\ref{subsec:EoS}, the
composition may change due to mixing of species within the flow.

Our computational domain spans a region of size $200 R_b \times 500
R_b$ (approximately $70\,$kpc$\,\times 175\,$kpc) in cylindrical
coordinates $(r, z)$ with an uniform grid whose resolution is 6 cells
per beam radius. This resolution is a compromise between a reasonable
computing time per model (about 185 hours on a NEC SX-5 vector
computer, running with a sustained performance of about 2.3 Gigaflops)
and the maximum evolutionary time to be reached in a simulation (see
Sect.\,\ref{subsec:caveats}).  Axisymmetry is assumed along the $r=0$
boundary, while the downwind boundaries at $r=200 R_b$ and $z=500 R_b$
are outflow boundaries. The boundary at $z=0$ is reflecting.

\section{Results} \label{sec:results}
\subsection{Morphology and dynamics}
\label{subsec:erg-morpho}
Due to the selected model parameters the beam radius and, hence, the
time units are slightly different for each model. They are given in
the first and second row of Table\,\ref{tab:modunits}.  The maximum
evolutionary time reached in each model is listed in rows three (in
code units) and four (in years), respectively. The last row gives the
evolutionary time (in code units) at which all models are compared.
This time corresponds to the final evolutionary time of the shortest
simulation (model LH) and is equal to $6.3 \cdot 10^6\,$y.
\begin{table}
 \caption{Length and time units for each model. The third and fourth
  row give the total duration of simulation $t_{max}$ in code and
  physical units, respectively. In the last row the time $t_{comp} 
  = 6.3 \cdot10^6\,$y is given in code units for each model. }
 \label{tab:modunits}
 \centering
 \begin{tabular}{@{}lccc}
  \hline
Jet  & BC & LC & LH \\
  \hline
$R_b$ [kpc]		& $0.366$	    & $0.361$	        & $0.342$ \\
$R_b/c$ [y]		& 1192 		    & 1176 	        & 1114 \\
$t_{max}$ [$R_b/c$]	& $5300$            & $5400$	        & $5950$ \\ 
$t_{max}$ [y]		& $6.3 \cdot 10^6$  & $6.4 \cdot 10^6$  & $6.6 \cdot 10^6$ \\  
$t_{comp}$[$R_b/c$]	& $5300$            & $5350$     	& $5650$ \\
  \hline
 \end{tabular}
\end{table}

Colour coded snapshots of the distributions of density, temperature
and Lorentz factor, nearly a the end of the simulations, are displayed
in Figs.\,\ref{fig:rho}--\ref{fig:W}.  Additional contour lines mark
the boundaries of the cocoon (see Sect.\,\ref{subsec:cocoon}). At
first glance, the morphology of all models is very similar,
particularly concerning the overall cavity and beam shapes.

Figure\,\ref{fig:profiles} displays profiles of several variables
along the symmetry axis. One notices that the differences among models
in quantities which are most important for the dynamics ($\rho$,
$\varepsilon$ and $P$) are small, at least far from the head of the
jet. The main differences occur where the profiles intersect the
biconical shocks and rarefactions within the beam which are quite
different (in spatial distribution and strength) from model to model.

The differences in the adiabatic index in the cavity (from model to
model) are the result of the different initial beam temperatures and
compositions. The leptonic models reach lower values of the adiabatic
index in the cavity than the baryonic one (i.e., they have $\gamma$
closer to $4/3$) because electrons and positrons become relativistic
at lower temperatures than protons. Although model LC has an average
temperature within the cavity (see Fig.\,\ref{fig:T}) more than ten
times smaller than in model BC, as a result of the mixing with the
ambient medium, the effective adiabatic index in the cavity is smaller
for model LC than for model BC. The reason being that model LC
supplies with a thousand times more relativistic particles the cavity
than model BC. The beam of model LH is the hottest and, thus, it has
the lowest adiabatic index. It is even lower than model LC because the
leptons of model LH are more relativistic (i.e., the beam temperature
is higher).

At the jet head different phases of vortex shedding can be seen in
Fig.\,\ref{fig:rho}. In model LC new vortices are forming, in model BC
they have been just shed from the head, and in model LH the Mach disk
has been replaced by a conical shock causing an acceleration of the
head. However, these differences are present only temporarily.  More
important are the differences inside the beam and in the cocoon which
we discuss next.
%
%

\subsection{Beam}
\label{subsec:beam}

In the hot, light model LH the density in the beam is two orders of
magnitude lower and the internal energy is, at least, two orders of
magnitude larger than in the two colder, denser models BC and
LC. However, model LH has a strongly pinched structure (at $z=39\,R_b$
and $z=63\,R_b$ in Fig.\,\ref{fig:W}) while the colder models possess
a continuous beam where the beam mass fraction is larger than 0.95
throughout the beam. The beam pinching, even disruption, is due to
considerable mass entrainment in model LH, inside which one can find
regions which are almost at rest. Hence, the light, hot jet appears to
be the most unstable one.

The thermodynamic quantities change drastically at the first conical
shock in model BC. The temperature rises from $10^{10}$K to
$10^{12}$K, the pressure (initially model BC is pressure matched, \ie
$K \approx 1$) increases by two orders of magnitude and stays at
values similar to those of the other models from there on ($K = {\cal
O}(100)$). The internal energy and sound speed also strongly increase,
while the relativistic Mach number drops from 130 to 20. All these
changes are caused by the initially very low internal energy of model
BC (see Tab.\,\ref{tab:models}). The other two models initially have a
much larger internal energy and the thermodynamic quantities in the
beam change less during the evolution. Actually, the properties of the
recollimation shocks are strongly varying from model to model. On
average, recollimations shocks have larger compression ratios and are
more numerous in model LH than in the colder models (see
Fig.\,\ref{fig:profiles}a).

The sound speed approaches its maximum value ($c_s \approx 0.57$)
close to the injection nozzle in model LH.  In the colder models $c_s$
increases within the beam and the maximum value is reached at the hot
spot. The deceleration (acceleration) at internal shocks
(rarefactions) within the beam is much more violent in model LH (where
Lorentz factors of up to 20 can be observed) than in the two colder
models.

At the beam/cocoon interface a thin, hot layer forms in the cold
models, which is thinner than that found in 3D simulations (Aloy \etal
1999, 2000). This layer results from the interaction of the beam with
the external medium, and appears naturally in some models of jet
formation \cite{SPA89}. Its existence has been proposed by different
authors \cite{Ko90,La96,Letal99} in order to account for a number of
observational characteristics of FRI radio sources.  However, the
physical nature of the shear layer is still largely unknown, as a
study of its properties requires much better resolution and full 3D
simulations. The structure of the shear layer is remarkably different
in model LH, where the hot layer forms out of the beam/cocoon
interface due to the extremely high specific internal energy of this
model. Nevertheless, the variations of pressure and Lorentz factor
across the shear layer are too small in model LH to have an effect on
the emission. This is in contrast to Aloy \etal\shortcite{Aletal00},
where the shear layer was much more extended and its emission
properties were distinguishable.

Despite of the differences between models, all of them maintain
relativistic beams up to distances of more than 80 kpc the average
Lorentz factor being larger than 5. This has important theoretical
consequences, in particular, for the most commonly accepted emission
models of kiloparsec-scale X-ray jets as we will discuss in
Sect~\ref{sec:diskussion}.

\subsection{Cocoon}
\label{subsec:cocoon}

The cocoon surrounding the beam is formed by beam matter that flows
backward (relative to the beam) after being deflected at the hot
spot. This beam matter partially mixes with the external medium, \ie
the cocoon does not contain pure beam material ($X \neq 1$). Hence, we
define the cocoon as the region containing material with a beam
particle fraction $0.1 < X < 0.9$. We have used the beam mass fraction
as the criterion, because the density within the cavity is similar in
all models (see Fig.\,\ref{fig:rho}), \ie the lower the value of $X$,
the lower is the number of emitting particles and thus the total
emitted intensity.  The thresholds are a bit arbitrary, but we have
checked that variations of the lower threshold (up to a decade) do not
change significantly the shape of the cocoon. For a very small lower
threshold ($X_{min}=0.001$) the cocoon practically coincides with the
complete cavity blown by the jet (see below). Increasing the value of
the upper threshold would include too much of the beam itself as part
of the cocoon. Other cocoon definitions have been given in, \eg Cioffi
\& Blondin \shortcite{CB92}.

According to our definition, the shape of the cocoon is remarkably
different among the considered models. In model LH the cocoon is
restricted to a narrow layer around the beam, while it is much more
extended and matches the relativistic backflow region in the
denser models (BC and LC). This is due to the lower beam density of
model LH, which allows for a larger inertial confinement of the beam
flow by the external medium without causing much mass entrainment
(which would destroy the jet). Another consequence of the lower
density of model LH is a lower beam mass fraction within the cavity
outside the cocoon (about two orders of magnitude smaller than in the
other models). The average cocoon temperature of models BC and LH is
around $10^{11}$K the size of local fluctuations reaching up to two
orders of magnitude. In model LC the average cocoon temperature is
lower ($10^{10}$K) and there are almost no fluctuations (see
Fig.\,\ref{fig:T}). The temperature distribution of models BC and LH
is very similar (although they have a different density and
composition), whereas that of model LC (with the same density as model
BC and the same composition as model LH) is very different.  In
Sect.\,\ref{sec:diskussion} we will explain how the varying particle
density causes this effect.

A backflow with mildly relativistic velocities occurs in thin shells
starting from the head of the jet and limiting the cocoon in models BC
and LH. The amount of beam material in this backflow is smaller in
model LC than in the other models (see, \eg Fig.\,\ref{fig:W}). 
%
%
%
%
%
%
%

\section{Evolution}
\label{sec:erg-evo}

\subsection{Definitions}

For the discussion of the jet evolution we introduce the following
quantities: (1) the jet length $l_j$ being the z-coordinate of the
contact discontinuity; (2) the cocoon length $l_{cc}$ being the
difference between $l_j$ and the smallest z-coordinate of the fluid
belonging to the cocoon (according to the definition given in
Sect.~\ref{subsec:cocoon}); (3) the cavity length $l_c$ being the
z-coordinate of the bow shock on the jet axis; (4) the average cocoon
radius $r_j$; (5) the average cavity radius $r_c$; (6) the cocoon
aspect ratio $A_j = l_{cc}/r_j$; (7) the cavity aspect ratio $A_c =
l_c/r_c$; (8) the hotspot pressure $P_{hs}$ being the average
pressure in the 10 cells upstream of $z=l_j$; (9) the average cavity
pressure $P_c$; (10) the jet velocity $v_j$ being the velocity of the
contact discontinuity between the jet and the external medium; and
(11) the bow shock velocity $v_c$ on the jet axis.

The position of the contact discontinuity ($l_j$) is always near the
bow shock ($l_c$, see Fig.\,\ref{fig:logevolution}a). But whereas
$l_c(t)$ is increasing monotonically, $l_j(t)$ looks rather `noisy'
due to the vortex shedding mechanism, which makes it difficult to
detect (numerically) the position of the contact discontinuity.  The
detection algorithm searches for the first cell on the jet axis
(beginning at $z=0$) where $X<0.5$. When a vortex sheds off, the
distance of this cell from the inlet is abruptly reduced.  As $A_j$
and $v_j$ depend on $l_j$, they display an oscillatory behaviour, too
(Figs.\,\ref{fig:logevolution}b and \ref{fig:logevolution}d,
respectively).  Other quantities like $r_j$ and $P_{hs}$ do also
oscillate because of the highly dynamical processes that affect them
(vortex shedding and mixing with the external medium), in particular
when the jets enter their second evolutionary phase (see
Sect.\,\ref{sec:evol-phases}).  $P_{hs}$ and $v_j$ have been smoothed
(averaged over 25 and 50 cells, respectively) in
Figs.\,\ref{fig:logevolution}a and \ref{fig:logevolution}c, but their
almost constant values during the first 100 time units are not caused
by this smoothing process.

\subsection{Evolutionary phases}
\label{sec:evol-phases}
%
%
%
%
%
%

In all models two evolutionary epochs can be distinguished.  During
the first epoch ($t < t^{\rm 1d} \approx 1.2\cdot10^5$y; see
Fig.\,\ref{fig:logevolution}) multidimensional effects are not yet
important and the jet propagates with a velocity close to the 1D
estimate \eqref{equ:v1d}. The Mach disk is only slightly disturbed,
the shedding of vortices is negligible, and both the hotspot pressure
and the jet velocity are nearly constant
(Figs.\,\ref{fig:logevolution}c, \ref{fig:logevolution}d). The aspect
ratios of the cocoon and of the cavity (Fig.\,\ref{fig:logevolution}b)
increase with time, while the average cavity pressure decreases
(Fig.\,\ref{fig:logevolution}c).

The second epoch starts when the first larger vortices are produced
near the jet head. The beam cross section increases and the jet
decelerates progressively. The deceleration is repeatedly interrupted
by short acceleration phases caused by the temporary replacement of
the Mach disk by a conical shock, which occurs \eg in model BC at $t
\approx 400$, 1200 and 5100, respectively (Fig.\,\ref{fig:vevolution}).  
The velocity increase during the acceleration phases can exceed 100\%.

The acceleration phases also give rise to a higher hotspot
pressure. Such pressure can be estimated assuming that the pressure of
the external medium is negligible.  Then the ram pressure of the
external medium $\rho_m v_c^2$ is equal to the thermal pressure behind
the bow shock, and hence about equal to the hotspot pressure as the
pressure is continuous across the contact discontinuity
(Fig.\,\ref{fig:pevolution}).  There is a phase shift between $P_{hs}$
and $\rho_m v_c^2$, because pressure fluctuations are generated at the
contact discontinuity (vortex shedding) which arrive at the bow shock
later.
\begin{table}
 \caption{Cavity lengths (in kpc) and aspect ratios at the end of the
one-dimensional epoch ($t = 1.1 \cdot 10^5\,$y; first two rows) and near
the end of the simulations ($t=6.0 \cdot 10^6\,$y; third and fourth
rows) for all models. The last row shows the time averaged velocity
during the time interval $[1.1 \cdot 10^5, 6 \cdot 10^6]$ (in units of
$c$).}
\label{tab:1d_vs_2d}
  \centering
   \begin{tabular}{@{}llccc}
   \hline
Phase &  Jet           &   BC    &   LC    &   LH \\
\hline
1D    & $l_c$          & $7.57$  & $7.10$  & $6.83$  \\
      & $A_c$          & $3.37$  & $3.09$  & $2.89$  \\
\hline
      & $l_c$          & $106.8$ & $100.6$ & $97.4$  \\
2D    & $A_c$          & $4.32$  & $4.03$  & $3.79$  \\
      & $\overline{v}$ & $0.055$ & $0.052$ & $0.050$ \\
\hline
 \end{tabular}
\end{table}

\section{Discussion}
\label{sec:diskussion}

In spite of an extremely different composition and internal energy,
the differences in morphology and dynamics of the jets are small
(particularly, between models BC and LC, both having the same density
contrast).  This is unexpected, as the jets of our models have a fixed
power which is carried by particles of different mass, \ie the
energies per particle are quite different.  A leptonic jet has
$\approx 10^3$ times more particles than a baryonic jet of the same
mass density. Therefore, its kinetic energy per particle is smaller,
\ie its temperature is smaller, but its specific internal energy
(erg/g) is larger. We also expected that differences in the adiabatic
index of the models would manifest itself.

It turns out, however, that fixing the power and initial speed of the
jet as well as the properties of the ambient medium, the development
of the jet seems to be quite well defined (at least for the jet
parameters considered and for the evolution time covered by our
simulations).  However, a more careful inspection of the results
reveals that some remarkable differences exist between our models
which we discuss in the following.

\subsection{Comparison with the extended Begelman-Cioffi model}
\label{subsec:BC89}

A simple theory of the evolution of the cocoon of extragalactic jets
was presented in BC89. The authors argue that the cocoon has not yet
reached pressure equilibrium with the surrounding medium in many
sources, and that high pressure confines the jet keeping it highly
collimated as seen in extragalactic sources. BC89 assume that both the
bow shock velocity $v_c$ and the power transported into the cavity
$L_c$ are not time-dependent, and that the pressure of the external
medium is negligible. Then the average cavity pressure is given by
\begin{equation}
        \label{equ:BC1}
        P_c = \frac{(\gamma_c-1)L_c}{v_c F_c} \, ,
\end{equation}
where $F_c = \pi r_c^2$ is the cross section of the cavity and
$\gamma_c=c_p/c_v$ is the constant average adiabatic index in the
cavity. The cavity pressure causes an expansion of the cavity with a
velocity $\dot{r}_c$, \ie
\begin{equation}
        \label{equ:BC2}
        P_c = \rho_m \dot{r}_c^2 \, .
\end{equation}
This implies $1/r_c \propto \dot{r}_c$ and hence
\begin{equation}
	r_c \propto t^{1/2}, \: F_c \propto t, \: P_c \propto t^{-1},
	\: l_j/r_c \propto t^{1/2} \, .
\end{equation}

We have extended the model of BC89 (eBC, hereafter) replacing the
assumption $v_c=const$ by the more realistic one $v_c \propto
t^\alpha$ and further assume that $\dot{r}_c \propto t^\beta$. Then
\eqref{equ:BC1} together with \eqref{equ:BC2} yields
\begin{equation}
        \frac{1}{t^{\alpha} t^{2(\beta+1)}} \propto t^{2\beta} \quad
        \Rightarrow \quad \beta = -1/2 -\alpha/4 \, ,
\end{equation}
and thus
\begin{equation}
\label{equ:modBCexp}
        r_c \propto  t^{\frac{1}{2}-\frac{\alpha}{4}}, \:\:
        F_c \propto t^{1-\frac{\alpha}{2}}, \:\: 
        P_c \propto t^{-1-\frac{\alpha}{2}}, \:\:
        l_j/r_c  \propto  t^{\frac{1}{2}+\frac{5\alpha}{4}} \, .
\end{equation}
The only free parameter of the eBC model is the value of
$\alpha$. A deceleration ($\alpha < 0$) leads to a faster radial
expansion, a slower decrease of the cocoon pressure and to a slower
increase of the aspect ratio. For $\alpha = -2/5$ the cavity evolves
self-similarly (\ie with constant aspect ratio).

\begin{table}
 \caption{Comparison of the simulation results with the extended BC89
  model.  Columns one and two give the evolutionary epoch and the
  model, respectively, and column three shows the value of the free
  parameter $\alpha$ obtained from the simulations by fitting the
  position of the bow shock as a function of time with a power
  law. The fourth column provides the fitted exponential time
  dependence of the cavity pressure, while the fifth column gives
  corresponding exponent of the eBC model ($= -1 -\alpha/2$).  Columns
  6 and 7 show the exponent describing the time dependence of the
  aspect ratio obtained from the simulations and the eBC
  model($1/2+5\alpha/4$), respectively.}
  \label{tab:exponents}
 \centering
 \begin{tabular}{@{}lcccccc}
\hline
Phase & Jet& $\alpha$ & $P_c$    &$P_c^{\rm eBC}$ &$l_j/r_c$& $(l_j/r_c)^{\rm eBC}$ \\
\hline
      & BC & $-0.113$ & $-0.945$ &       $-0.944$ &   $0.449$ &  $0.358$ \\
1D    & LC & $-0.151$ & $-0.940$ &       $-0.925$ &   $0.404$ &  $0.311$ \\
      & LH & $-0.219$ & $-0.906$ &       $-0.891$ &   $0.358$ &  $0.226$ \\
\hline
      & BC & $-0.355$ & $-0.722$ &       $-0.823$ &   $0.058$ &  $0.056$ \\
2D    & LC & $-0.356$ & $-0.737$ &       $-0.822$ &   $0.071$ &  $0.054$ \\
      & LH & $-0.363$ & $-0.779$ &       $-0.818$ &   $0.059$ &  $0.045$ \\
\hline
 \end{tabular}
\end{table}

We have extracted the parameter $\alpha$ and the exponential
dependence of $P_c$ and $l_j/r_c$ from our simulation results by
fitting the position of the bow shock as a function of time with a
power law (Tab.\,\ref{tab:exponents}). Using the fitted values of
$\alpha$ we have also computed the exponents for $P_c^{\rm eBC}$ and
$(l_j/r_c)^{\rm eBC}$ according to the extended BC89 model.  The
fitting procedure is not very reliable for the 1D epoch as it involves
only a few data points, while velocity fluctuations cause problems for
fitting the 2D epoch. Thus, differences between the models are
probably not significant. Note in this respect that although the value
of $\alpha$ varies by a factor of two between the models, the
variation of $P_c^{\rm eBC}$ is at most 5\% and that of
$(l_j/r_c)^{\rm eBC}$ less than 25\% as both quantities can be
extracted more reliably.  The values of $P_c^{\rm eBC}$ and
$(l_j/r_c)^{\rm eBC}$ obtained from the simulations and from $\alpha$
via the eBC model agree reasonably well. The exponent for the aspect
ratio is slightly larger than that predicted by the eBC model
(especially for the 1D epoch), \ie the radial expansion of the cavity
is slower than expected from the deceleration of the jet head. This
discrepancy arises from the assumption of the BC89 model that the
whole power of the jet is used to increase the pressure behind the bow
shock (which is responsible for the sideways expansion of the
cavity). However, part of the jet power is used to fuel the
relativistic backflow and its kinetic energy is not available to
expand the cavity in radial direction, \ie the aspect ratio (jet
length over cavity radius) grows faster in the simulations than in the
BC89 model.

During the 2D epoch the cavity evolution is almost self-similar and
the cavity aspect ratio remains nearly constant.  This situation will
only change when the cavity pressure becomes equal to the pressure of
the external medium and the radial expansion ends. Extrapolating from
Fig.\,\ref{fig:logevolution} this will happen after $\sim 10^6$ time
units or $10^9$\,y, which exceeds the estimated ages of observed
jets ($10^7$\,y -- $10^8$\,y) by one order of magnitude.

The evolution of the jets proceeds in two epochs independent of the
composition of the jet. This is in agreement with previous results
obtained by MMI98 and Komissarov \& Falle \shortcite{KF98}. The 1D
epoch of our models corresponds to the intermediate phase of
Komissarov \& Falle \shortcite{KF98}, while the second epoch may be
identified with their self-similar phase. They also find that the
growth rate of the aspect ratio is smaller during the second
epoch. However, their growth rates are slightly larger than ours in
both epochs. This discrepancy is probably due to different jet
parameters. Their relativistic models are 30 times denser than ours,
have a little bit smaller Lorentz factor, are pressure matched, and
possess a finite opening angle. The transition to the second epoch
does not occur in those models of Komissarov \& Falle \shortcite{KF98}
which have an opening angle larger than $10^\circ$.

\subsection{Evolutionary differences}
\label{subsec:evoldiff}

The average jet velocities during the 2D epoch are listed in
Tab.\,\ref{tab:1d_vs_2d} together with the lengths and the cavity
aspect ratios at the end of the 1D epoch and at the end of the
simulation. At first glance there seems to be a clear trend in the 1D
epoch: with increasing internal energy and decreasing Mach number the
jets become slower and wider (the mass flux into the cocoon
increases). As already found by MMI98, the jet velocities are
noticeably larger than predicted by the 1D estimates ($v_j^{\rm 1d} =
0.2c$) at the beginning of the simulation (see
Figs.\,\ref{fig:logevolution}d and \ref{fig:vevolution}).  This
discrepancy is caused by imposing a reflecting boundary condition at
$z=0$ (see Sect.\,\ref{subsec:models}) and by the different pressure
ratios $K$.  After some ten time units the jet head has propagated far
enough from the grid boundary to be no longer affected by it. The
average cavity pressure decreases and the pressure contrast between
the jet and the cavity becomes closer to one in all the models
(Fig.\,\ref{fig:logevolution}).  After the jets has decelerated at the
end of the 1D epoch their velocities and the 1D estimates agree very
well (Fig.\,\ref{fig:vevolution}).

During the 2D epoch the jet of model BC propagates considerably faster
than those of models LC and LH, while the jet of model LC is only
slightly faster than that of model LH.  This result can be easily
understood.  The relativistic Mach number of model BC is larger, \ie
the angle between the conical shocks inside the beam and the beam axis
is smaller. Hence, these shocks are less efficient in decelerating the
flow. This also affects the propagation speed of the head, because the
terminal Mach disk is temporarily replaced by conical shocks an effect
pointed out already by Mart\'{\i} \etal (1997). The noisy bow shock
velocity (Fig.\,\ref{fig:vevolution}) is also a result of this
process. However, as the heads of all jet models are hot and the Mach
number is low in this region, this effect does not lead to very strong
differences between the computed models.
  
The differences in the cavity aspect ratios correlate with the average
propagation speeds. The jet of model BC is the fastest and the longest
one (Tab.\,\ref{tab:1d_vs_2d}). The cavity radius is similar for the
cold models and bigger for model LH. The models expand radially almost
at the same rate, because the average cavity pressure is essentially
equal for all three models although a significant part of the kinetic
energy which is contained in the relativistic backflow cannot be used
to inflate the cavity.  In fact, we included model LH in our
investigation in order to check whether an increased fraction of
internal (non-directed) energy would lead to less kinetic (directed)
energy in the cocoon. 
%

\subsection{The influence of different compositions}
\label{subsec:composition}

From a numerical point of view the most important effect of a varying
composition is the non-constant adiabatic index requiring an extension
of the original MMI98 code (see Sect.\,\ref{subsec:models}). The
composition dependent $\gamma$ calculated from the Synge EoS (see
Sect.\,\ref{subsec:EoS}) however, does not lead to large differences
in the evolution and morphology of the jets. The average $\gamma$ in
the cocoon is around $1.4$ and $\gamma=5/3$ in the external medium for
all three models. Due to the initial conditions, the adiabatic index
within the beam is different for every model, but these differences
have no observational relevance as $\gamma$ is not an observable
quantity. 

We expected that the largest internal energy content in the leptonic
models would cause a larger expansion of the beam than in the baryonic
model, leading to a faster deceleration of the head and differences
in the aspect ratio of the cavity (more spherical in the leptonic
models). However, this effect is compensated by the large overpressure
of the cavity that efficiently confines the jet laterally.

The variation of the particle density with the composition gives rise
to huge differences in the temperature. At the same density, an
$e^\pm$-gas contains three orders of magnitude more particles than an
ionised hydrogen gas. The particle density is proportional to $\eta
\cdot X_{lb}$ and, therefore, it is much higher in model LC than in
the other two models. Hence, the same amount of energy is distributed
over many more beam particles in model LC than in models BC and LH,
leading to an about three orders of magnitude lower beam temperature
(Fig.\,\ref{fig:T}). However, this does neither explain the flat
temperature profile of model LC, nor the identical average cavity
temperatures of models LH and BC (whose hotspot temperatures differ
by a factor of ten). To explain the temperature differences in the
cocoon, particles from the external medium must be considered
too. There are $N_{cm}=7 \cdot 10^{64}$ particles from the external
medium and $N_{cb}=3 \cdot 10^{65}$ particles from the beam in the
cavity of model LC at $t=6.0 \cdot 10^6$y. The dominant fraction of
the internal energy of the cavity is also carried by beam
particles. Hence, the mixing of beam and external medium has little
impact on the internal energy per particle, \ie the temperature
remains at the hotspot value.

Concerning models LH and BC, particles from the external medium
dominate the particle density in the region of the cavity excluding
the dense shell behind the bow shock ({\it cavity interior}), \ie
$N_{c_ib}/N_{c_im} < 1$, but the internal energy is still dominated by
beam particles. This leads to large changes and fluctuations of the
temperature during the mixing of particles from the beam and the
external medium (Fig.\,\ref{fig:T}). The temperatures near the jet
head and near the inlet are very different in these two
models. Although the hotspot temperatures differ, the average cocoon
temperatures are similar because the same amount of energy is
distributed over a similar number of particles. In addition, since the
particle density is much lower than in model LC, the average cocoon
temperature is much higher. 

In general, when $N_{c_ib}/N_{c_im}>1$ the cavity interior has a flat
temperature profile and we talk of an {\it isothermal} cavity.  When
$N_{c_ib}/N_{c_im}<1$ the average temperature in the cavity interior
is heterogeneous and independent of the beam composition and
density. In this case we talk of a {\it non-isothermal} cavity.
Another effect is caused by the different time dependencies of
$N_{c_ib}$ and $N_{c_im}$. The number of particles from the external
medium in the cavity interior is proportional to its volume, and the
cavity interior has a roughly constant density during the whole
evolution (\eg, it changes by less than 20\% during $6.0 \cdot 10^6$y
in case of model LH). Hence, from the eBC model one can derive
$N_{c_im} \propto t^{2(1+\beta)}\cdot t^{1+\alpha} =
t^{2+\alpha/2}$. While the number of beam particles increases in the
cavity interior linearly with time ($N_{c_ib} \propto t$), the ratio
$N_{c_ib}/N_{c_im} \propto t^{-1-\alpha/2}$ decreases, \ie a
transition from an isothermal to a non-isothermal cavity may occur
during the jet evolution.
%
%

Figure\,\ref{fig:nratio} shows the evolution of $N_{c_ib}/N_{c_im}$ for
the three models. Model LC has an isothermal cavity throughout the
simulation, while the cavities of the other two models are always
non-isothermal. Although, we have not found the above mentioned
transition, Fig.\,\ref{fig:nratio} suggests that one should be able to
find initial jet parameters (between those of models LH and LC) where
such a transition will occur during the simulation.

In an isothermal cavity the external particles can be neglected, \ie
$N_c \approx N_{c_ib} \propto t$. As the cavity volume is proportional
to $t^{2+\alpha/2}$, the number density within the cavity $n_c \propto
t^{-1-\alpha/2}$. However, as $P_c \propto t^{-1-\alpha/2}$, this
implies that the temperature $T_c \propto P_c/n_c$ does not depend on
time. This explains why the cocoon temperature is constant in
$z$-direction and also in $r$-direction (Fig.\,\ref{fig:T}). In a
non-isothermal cavity particles from the external medium dominate, \ie
$n_c = const$ and the average cocoon temperature decreases according
to $T_c \propto t^{-1-\alpha/2}$.

\subsection{Comparison with observations}
\label{subsec:observations}

 Comparing the hydrodynamic properties at the end of our simulations
with those deduced from observations of Cyg\,A, we find that the hot
spot pressure ($P_{hs} \approx 0.001\rho_mc^2 \approx 1.5 \cdot
10^{-9}\,$dyn cm$^{-2}$) agrees within a factor of 2 with the value
reported by Carilli \etal \shortcite{Cetal96}.  Comparing the hotspot
advance speeds of CSO and FR II sources implies a deceleration of the
hotspot propagation at early epochs. This deceleration indeed occurs
in our simulations.  The propagation velocity after the deceleration
($v_j \approx 0.05c$) is in good agreement with the estimates of Daly
\shortcite{Da95}. The equipartition magnetic field of $\simeq 60\,
\mu$G in the cavity and $\simeq 600\,\mu$G in the hotspot are larger
than those reported for Cyg~A \cite{Cetal96}, although they agree with
the values obtained for the radio lobes and hotspots of younger
powerful radio sources \cite{Fe98}. This agreement is easily explained
considering that the evolutionary time covered by the simulations is
relatively short.

Up to now it is impossible to determine hydrodynamic properties of
jets directly from the observed radio emission. Thus, in order to
compare the simulated jets with observed ones we proceed in 'opposite'
direction and compute the non-thermal synchrotron radio emission of
our hydrodynamic models. We follow the same approach as in Aloy \etal
\shortcite{Aletal00} assuming that the energy density weighted with
the beam particle fraction is distributed among the emitting
non-thermal particles according to a power law $N(E) \propto
E^{-\sigma}$ (without cut-offs). The magnetic field is considered to
have equipartition strength and a {\it ad hoc} structure: aligned with
the flow velocity and with a negligible random component (because we
are not concerned in polarisation properties).  The simulated radio
maps obtained from the hydrodynamic models (Fig.\,\ref{fig:radiomap})
include only those computational zones where the density is smaller
than 0.1$\rho_m$ to exclude emission from the external medium. We have
chosen a viewing angle of $45^\circ$ to compute the synthetic
radio maps in order to avoid excessive Doppler boosting of the beam
which would out-shine completely the diffuse emission from the cocoon.
%
%

  Two main features known from observations are present in all models:
radio lobes with hotspots and a one-sided, knotty jet (particularly
evident in model LH). The knots are associated with internal shocks in
the beam. Emission is dominated by the hotspots, where the pressure
is maximum. Figure\,\ref{fig:radiomap} shows that the models have
different jet/cocoon emission ratios, as expected from their different
cocoon morphology (Sect.\,\ref{subsec:cocoon}).  While models BC and
LC show some emission from the cocoon, resembling the observed lobes,
the emission is dominated by the beam in model LH.

 In actual sources the lobes of the jet and counter-jet are often
observed to be equally prominent. Our simulations show larger lobes
(\ie more cocoon emission) for the counter-jet (particularly for models
BC and LC). This can be explained by the relativistic backflow present
in the simulated models, because Doppler boosting enhances the
emission from the backflow regions in the counter-jet and dims that of
the approaching jet lobe. The relativistic backflow also limits the
radial expansion of the lobes and prevents the beam from inflating
large lobes with gas moving at sub-relativistic speeds. Therefore, the
similarity of the lobes in actual radio sources suggests that there
are no relativistic backflow regions. Their presence in the simulated
models is most likely due to the axial symmetry imposed in our 2D
simulations. Three dimensional simulations \cite{Aletal99} show much
smaller backflow velocities, typically $\sim 0.25c$, and hence do not
lead to substantial Doppler beaming in the counter-jet cocoon.

Very recently, the new generation of space-based X-ray observatories
has allowed for the detection of X-ray emission from kiloparsec scale
jets \cite{Schetal00,Chetal00,Saetal01}. The observed X-spectra are
not satisfactorily explained by standard radiation mechanisms, \ie
synchrotron self-Compton \cite{TMG98,Schetal00,Taetal00} and
synchrotron radiation.  The latter can arise from the same population
of particles that produce the radio emission
\cite{Schetal00,Maretal01}, or from a second, much more energetic
population of electrons co-spatial with the one responsible for the
radio emission \cite{Roetal00,Saetal01}.  Alternatively, inverse
Compton scattering of beamed photons of the cosmic microwave
background (CMB) has also been proposed \cite{Taetal00,Ceetal01}. For
this mechanism to work, it is necessary that the beam of the jet is,
at least, mildly relativistic, because the amplification of the CMB
radiation increases with the Doppler factor, $\delta \equiv [\Gamma (1
- \beta \cos \theta)]^{-1}$ \cite{Taetal00}. Our models show
relativistic beam velocities ($\Gamma \simeq 5$) out to 80 kpc (see
Sect.\,\ref{subsec:beam}).  Whether this is a result of the imposed
axisymmetry of our models or a physical feature may only be
disentangled by performing three-dimensional simulations (see
Sect\,\ref{subsec:caveats}).

Thermal bremsstrahlung from the gas confining the radio-optical jet
was very soon disregarded as the source of the observed X-ray
radiation from the jets of powerful sources \cite{Saetal01,Schetal00},
although the data can be fitted by a thermal bremsstrahlung spectrum
\cite{Chetal00}. In order to explain the observations invoking thermal
bremsstrahlung, a large electron density ($n_e \simeq 2\,$cm$^{-3}$)
is required. This is, however, inconsistent with the observed
(relatively low) rotation measure ($n_e < 3.7 \times 10^{-5}/(B\,L)$;
$B$ and $L$ are the magnetic field strength and the path length,
respectively; Schwarz \etal 2000).  The estimate for the electron
number density, is based on the assumption that (i) there are no
magnetic field reversals within the VLA beam, and (ii) the confining
thermal material is not relativistic and relatively cold. But, our
simulations show the presence of a very hot relativistic plasma (at
least in model BC), which is not accurately described by the classical
formulas.

Given that the densities are very similar in all our models and that
there are huge temperature differences inside the cavity
(Fig.\,\ref{fig:T}), we have analysed whether relativistic thermal
bremsstrahlung can be used to distinguish between the cavities
inflated by jets of different composition.  We follow the work of
Nozawa, Itoh \& Kohyama \shortcite{NIK98}, which includes the
appropriate Elwert \shortcite{El39} factor to compute the relativistic
thermal bremsstrahlung cross section.  Their method is accurate if one
can neglect the thermal motion of protons, \ie at temperatures below
$10^{11}\,$K. For larger temperatures the relativistic thermal
bremsstrahlung cross section is still accurate within an order of
magnitude (Itoh, private communication).  Above $10^9\,$K
electron-electron, electron-positron, positron-ion, and
positron-positron bremsstrahlung processes are important (see, \eg
Novikov \& Thorne 1973). These processes should have been included
when computing the total bremsstrahlung emissivity of our models,
where temperatures are as high as $10^{12} - 10^{13}\,$K in the shear
layer confining the jet, in the hotspot, and in the cocoon of model
BC. However, as the emissivity of these processes is comparable with
that produced by the interaction of electrons with protons
\cite{De86,Sv82}, the emissivity of the latter process provides a
lower bound of the total bremsstrahlung emissivity, which is accurate
to an order of magnitude up to temperatures of $\sim 10^{13}\,$K.

%
%

The bremsstrahlung power $P_{\nu}^{brems}$ per unit of frequency
($\nu$) in the comoving frame is proportional to $n_e n_i / \sqrt{T}$
($n_i$ is the number density of ions in the medium). Considering a
viewing angle of $90^{\circ}$, \ie beaming effects are unimportant,
$P_{\nu}^{brems}$ is dominated in all models at relatively low
frequencies ($h \nu \la 100\,$keV) by the very dense shell at the
cavity boundary behind the bow shock containing shocked external
medium.  Although the non-shocked external medium is cooler than the
shell, $P_{\nu}^{brems}$ is three orders of magnitude smaller than in
the shell, because the shocked material is much denser than the
ambient medium. The bremsstrahlung emission from the cavity's interior
is very weak in models BC and LH.  This is different for model LC,
where the temperatures in the cocoon are lower and the electron
densities are larger than in the other models. The resulting
bremsstrahlung emissivity of the cavity's interior is almost as strong
as that of the external medium.

The dominance of the X-ray emission from the shell is also evident
from the surface brightness, \ie from the integral of
$P_{\nu}^{brems}$ computed along the line of sight
(Fig.\,\ref{fig:bremss_90_bk}). The total emission X-ray maps of all
models look very similar, as the shell outshines the cavity at a
frequency of $10\,$keV (Fig.\,\ref{fig:bremss_90_bk} top panel).  Note
that our analysis does not take into account the emission of
intra-cluster gas, which would be dominant for a radio source located
in a galaxy cluster. The dominance of the shell is in agreement with
the work of Heinz, Reynolds \& Begelman \shortcite{HRB98}. They also
find that the shell's dominance decreases as the source evolves
because the bow shock decreases in strength when the pressure inside
the cavity decreases due to the expansion.  McNamara \etal
\shortcite{Mcetal00} find a reduced X-ray surface brightness when
observing the radio lobes of Hydra\,A, an effect which is evident in
our models. However, McNamara \etal \shortcite{Mcetal00} also point
out that there is no evidence for shock-heated gas surrounding the
radio lobes, and hence suggest that the cavity expands subsonically.
This does not contradict our results, as we consider FR\,II sources
(Hydra\,A is a FR\,I source) and because our models represent a still
relatively young evolutionary stage of a powerful radio source.  FR\,I
sources reach the transonic regime relatively soon and, therefore, one
does not expect to find a strong shock surrounding the radio lobes.
Instead the radio lobes flare into the external medium. This may
explain the lack of X-ray emission from the region surrounding the
cavity. On the other hand, it is expected that a further evolution of
the source will lead to a decrease of the cavity pressure, and
consequently to a weakening of the cavity bow shock (which in turn
reduces the X-ray emission of the shell).

We have also computed the surface brightness per unit of frequency at
$10\,$MeV (Fig.\,\ref{fig:bremss_90_bk} bottom panel). We choose this
high frequency, which is beyond the observational range of {\it
CHANDRA} ($\approx 0.1 - 10\,$keV), because only when considering hard
X-rays one begins to see the emitting fluid inside the cavity and the
emission of the model differs most. Increasing the frequency from
$10\,$keV to $100\,$keV$\, \la \nu \la 1\,$MeV the interior of the
cavity shows up, although the bow shock surrounding the head of the
jet is still dominant. At even higher frequencies ($\ga 1\,$MeV), the
hotspot and the high temperature cocoon are visible.  Finally, at $h
\nu = 10\,$MeV (bottom panel in Fig.\,\ref{fig:bremss_90_bk}), only
models BC and LH show some emission while the exponential cut-off of
the bremsstrahlung spectrum reduces the emissivity of model LC, as the
temperature of the cavity interior of model LC, $kT \simeq 500\,$keV,
is much smaller than the considered frequency. This might provide a
way to distinguish observationally -- once future detectors have the
appropriate dynamical range and sensitivity -- leptonic from baryonic
jets (with the same kinetic power and density contrast).  If one
observes a transition from a shell dominated source to a source
dominated by the interior of the cavity when increasing the
observation frequency from $h \nu \approx 1\,$MeV to $h \nu \approx
10\,$MeV, the jet feeding the radio lobe is most probably rich in
baryons.  Conversely, if this transition is not detected and one finds
instead a significant suppression of the emission, the jet is mostly
made of leptons.

Typical luminosities of very powerful X-ray, large-scale jets are of
the order $10^{45}\,$erg$\,$sec$^{-1}$, \eg Chartas \etal
\shortcite{Chetal00} report a X-ray luminosity in the $2 - 10\,$keV
band of $L_X \simeq 6.3 \cdot 10^{45}\,$erg$\,$sec$^{-1}$ for PKS
0637-752. The X-ray luminosities produced by our jet models due to
thermal bremsstrahlung are far below the observed values.  This rules
out thermal bremsstrahlung as the dominant mechanism for the X-ray
emission of large-scale jets. However, several factors may enhance
the total thermal luminosity. First, the size of our jets is still
relatively small as compared with fully evolved FR II sources (the
larger the jet the larger is the luminosity). Second, the kinetic
power of our models, although being large, is still below the one
inferred \eg for PKS 0637-752 ($10^{48}\,$erg$\,$sec$^{-1}$; Tavecchio
\etal 2000).  Our simulations show that the jet itself radiates only a
small fraction of the total power per unit of frequency emitted by the
whole cavity ($1.5 \cdot 10^{22}\,$erg$\,$sec$^{-1}$Hz$^{-1}$ for
model BC at $10\,$keV). At $10\,$MeV the emitted power from the cavity
is $2.7 \cdot 10^{16}\,$erg$\,$sec$^{-1}$Hz$^{-1}$ and is dominated by
the hotspot which contributes $8.0 \cdot
10^{14}\,$erg$\,$sec$^{-1}$Hz$^{-1}$.

\subsection{Caveats}
\label{subsec:caveats}

Which are the astrophysical and numerical limitations of our
simulations? 

We have used a uniform external atmosphere, but extragalactic jets
after having crossed a galactic halo will propagate into a much more
diffuse intergalactic (or intra-cluster) medium. Both the pressure and
the density decrease with the distance from the galactic nucleus. A
density declining atmosphere will cause a widening of the jet (and of
the cavity), and a substantial deceleration provided that the jets are
not inertially confined. The effects of (continuous or abrupt) density
changes in the external medium and its impact on the long term
evolution of Newtonian jets has been investigated, \eg by Hooda \etal
\cite{HMW94}. Such simulations involve additional, not very well
determined model parameters and are more complex making them less
attractive from the computational point of view, in particular as our
grid resolution (zones per beam radius) is much larger than that in
Hooda \etal \shortcite{HMW94}.

Our simulations are purely hydrodynamic, \ie they do not consider the
effects of magnetic fields. Strong magnetic fields might confine the
shocked jet material in an extended {\it nose cone} preventing it to
be continuously deposited into the cocoon. An episodic release of
thermal material in the cocoon would have important consequences for
the stability of the jet. However, it is quite unlikely that at kiloparsec
scales strong magnetic fields do exist. Instead, jets are expected to
transport weak and mainly randomly oriented magnetic fields (\eg
Ferrari 1998).

%
%

The grid resolution employed (six zones per beam radius at the
injection nozzle) is far from being appropriate to account for
phenomena like mass entrainment from the external medium. However, it
is sufficient to describe the gross morphological features and to
capture the average cocoon dynamics. One should consider that, in
order to compute up to a maximum physical time, $t_{end}$, the
required run time is proportional to $(\rm{cells}/R_b)^3$. Hence,
increasing the grid resolution by a factor of two requires about 1200
hours of computing time to reach $t_{end}\simeq 6 \cdot 10^6\,$y, and
9600 hours to simulate a typical lifetime of a FR\,II source
($t_{end}\simeq 10^7\,$y).  In order to check whether the chosen
resolution was sufficient to capture, at least the qualitative
morphology, we performed a set of shorter simulations $t_{end} \simeq
2.3 \cdot 10^4\,$y at different grid resolutions
(Fig.\,\ref{fig:resdep}).  With four zones$/R_b$ the short
acceleration phases caused by the vortex shedding are missing, while
they are present with six zones$/R_b$. At even higher resolution the
qualitative changes are small.

Finally, we have restricted ourselves to axisymmetric jet flows. This
is a considerable restriction, because we force the jet to propagate
along the symmetry axis. As, \eg Aloy \etal \shortcite{Aletal00}
showed, three-dimensional jets may wobble and their effective head
area may change appreciably. Consequently, two possibilities arise:
(i) the effective area grows (as the result of, \eg the {\it dentist
drill effect}; Cox, Gull \& Scheuer \shortcite{CGS91}) leading to a
faster deceleration , or (ii) the wobbling of the Mach disk reduces
the cross sectional surface of the jet head leading to an acceleration
\cite{Aletal00}. It is difficult to extrapolate the results of Aloy
\etal \shortcite{Aletal00} and to decide whether or not such
mechanisms operate in the long run, and what their consequences are.
Thus, three-dimensional numerical simulations are required to clarify
this issue.

Because of the assumed axisymmetry we encountered numerical problems
when simulating under-pressured jets (see Sect.\,\ref{subsec:param}).
These problems would not appear in three--dimensional simulations as
the jet fluid will bypass the blobs of external medium material trying
to block the inlet.

\section{Summary and Conclusions}
\label{sec:conclusions}

 We have performed the longest and best resolved large-scale
simulations of relativistic jets up to date. The simulations extend up
to an evolutionary age of $6 \cdot 10^6\,$y which is only a few times
smaller than the ages of typical powerful radio sources. The simulated
jets differ in their density, temperature and composition, but have
similar gross properties (kinetic luminosity and thrust). Hence, they
may represent a relatively early stage of the observed extragalactic
jets and their radio lobes. We have considered three models including a
hot and a cold jet made of a pure (leptonic) electron-positron plasma
(models LH and LC) and one baryonic model BC consisting of an
electron-proton plasma.  The leptonic hot model LH has a factor of 100
lower density ($\eta=10^{-5}$) than the other two models
($\eta=10^{-3}$).

 Although there is a difference of about three orders of magnitude in
the temperatures of the cavities inflated by the simulated jets, we
find that both the morphology and the dynamic behaviour are almost
independent on the assumed composition of the jets. Only the leptonic
hot model LH behaves differently because of its very light beam. This
reflects the well known influence of the density parameter $\eta$
according to which light jets are less stable than heavy ones. Model
LH suffers from considerable mass entrainment into the beam, which
eventually leads to the disruption of the beam. The instability grows
on a time scale of a few million years, \ie the evolution of model LH
is inconsistent with the sizes and ages of the jets of typical
powerful radio sources. In addition, the initial conditions chosen for
model LH are quite extreme, and most likely cannot be found in actual
jets.

 In all models two evolutionary epochs can be distinguished.  During
the first epoch multidimensional effects are not yet important and the
jet propagates with a velocity close to the 1D estimate. The Mach disk
is only slightly disturbed, the shedding of vortices is negligible,
and both the hotspot pressure and the jet velocity are nearly
constant.  The aspect ratios of the cocoon and of the cavity increase
with time, while the average cavity pressure decreases.  The second
epoch starts when the first larger vortices are produced near the jet
head. The beam cross section increases and the jet decelerates
progressively. The deceleration is repeatedly interrupted by short
acceleration phases caused by the temporary replacement of the Mach
disk by a conical shock. The velocity increase during the acceleration
phases can exceed 100\%. 

 The evolution of the cocoon and cavity aspect ratios of all models is
in agreement with an extended version of the simple theoretical model
proposed by Begelmann \& Cioffi (BC89) for the inflation of the
cocoon. The cavity aspect ratio of the simulated models is too large
according to current observations, \ie the simulated jets are too
elongated. This situation may change when the simulations are either
extended even further in time or when the assumption of axisymmetry is
relaxed.  However, both approaches would require prohibitive amounts
of computational resources and axisymmetry is, in general, a good
approximation to study extragalactic jets.

 The size and shape of the cocoon depends on $\eta$, \ie on the ratio
of the densities in the jet and in the ambient medium. According to
our definition of the cocoon -- the region containing material with a
beam particle fraction $0.1 < X < 0.9$ -- we find that a light jet
(model LH) has a very thin cocoon restricted to a narrow layer girding
the beam. Jets with heavier beams (models BC and LC) possess more
extended cocoons, more similar to those of observed jets.  The
differences in the cocoon sizes become particularly evident when
comparing the synthetic radio maps that we have computed for all
models.  The cocoon aspect ratio found for the leptonic hot model LH
is unrealistically large, which provides another argument against its
realization in nature (see above).

The propagation velocities and the hotspot pressures of our jets
agree very well with the analytical and observed values. The strength
of the equipartition magnetic field comfortably matches the
expectations for a young powerful radio source (whose magnetic fields
are thought to be stronger than those of evolved radio sources). The
beam velocities of the models are relativistic ($\Gamma \simeq 4$) at
kiloparsec scales. Provided this is not an artefact caused by the
assumed axisymmetry, the results strongly support the ideas of
Tavecchio \etal \shortcite{Taetal00} and Celotti \etal
\shortcite{Ceetal01} that the X-ray emission of several extragalactic
jets may be due to relativistically boosted CMB photons.

 The radio emission of all models is dominated by the contribution of
the hotspots. The two cold models (BC and LC) have very similar
radio-morphologies with a continuous beam connecting the inlet with
the hotspot. This is different from model LH where the synthetic
radio maps show a discontinuous and knotty beam even at a viewing
angle of $45^{\circ}$ (where the effects of Doppler boosting are still
small).

 We find that the thermal bremsstrahlung emission is strongly
suppressed at very high energies ($\approx 10\,$MeV) in the leptonic
cold model because the average temperature of its cavity ($\sim
10^{10}\,$K) is much lower than that of the two other models ($\sim
10^{11}\,$K). The baryonic and the leptonic hot model both show a
gradual transition in the characteristics of their thermal
bremsstrahlung emission. At low energies the emission is dominated by
the shell, while the interior of the cavity dominates at high
energies. All models exhibit a depression in the X-rays surface
brightness of the cavity interior in agreement with recent
observations \cite{Mcetal00}.  These observations also show a lack of
strong X-ray emission from the boundaries of the radio lobes of
Hydra\,A. This does not contradict our results as Hydra\,A is a FR\,I
source, while we have been simulating powerful radio sources. In a
powerful source the bow shock is a strong shock and, hence, it may
produce substantial X-ray emission. FR\,I sources flare into the
external medium and, most probably, they do not have strong bow
shocks.

\section*{Acknowledgments}

The authors gratefully acknowledge enlightening discussions with
Torsten Ensslin, Sebastian Heinz, Gopal-Krishna, Naoki Itoh and Jos\'e
M. Ib\'a\~nez.  M.A.A. and J.M.M. received support through an
agreement between the Max-Planck-Gesellschaft and the Consejo Superior
de Investigaciones Cient\'{\i}ficas.  M.A.A. also acknowledges the
EU-Commission for a fellowship (MCFI-2000-00504).  This research was
supported in part by the Spanish Direcci\'on General de Ense\~nanza
Superior (DGES) grants PB97-1164 and PB97-1432, and by the Fulbright
commission for collaboration between Spain and the United States.


\begin{figure*}
%
\caption{Values of $\eta$ (solid lines) and $T_b$ (dashed lines) for
leptonic jets ($X_{lb}=1$) as a function of $R_b$ and $W_b$ after
fixing $L_{kin}$, $v_j^{1d}$, $X_{lb}$ and the external medium
according to Tab.~\ref{tab:models}.  There are no physical solutions
in the areas without $T_b$-contour lines.  The thick, solid lines
correspond to values of $\eta$ (from top to bottom) of $0$, $10^{-4}$,
$10^{-3}$, $10^{-2}$.  The thick, dashed lines correspond to values of
$T_b$ (from left to right) of $10^{10}$K, $10^{9}$K and $10^{8}$K.
For baryonic jets the diagram would be similar, except for the values
at the thick, dashed lines which would be three orders of magnitude
higher: $10^{13}$K, $10^{12}$K and $10^{11}$K, respectively -- the
reason being that for the same value of $\eta$, one would have a value
of $X_{lb}$ three ten folds larger. The squares mark the situation of
the three simulated models in this diagram (see \ref{subsec:models}).}
\label{fig:pspace1}
\end{figure*}

\begin{figure*}
%
\caption{Snapshots of the logarithm of the density ($\log_{10}
(\rho/\rho_m)$) for the models BC (top panel), LC (central panel) and
LH (bottom panel) at $t \approx 6.3 \cdot 10^6$y. The black lines are
iso-contours of the beam mass fraction with $X = 0.1$ (outermost) and
$X = 0.9$ (innermost). These values correspond to the boundaries of
the cocoon and the beam, respectively. Please, note that the colored
version of the figure is only provided in the electronic version of
the journal.}
\label{fig:rho}
\end{figure*}

\begin{figure*}
%
\caption{Snapshots of the logarithm of the temperature
($\log_{10}(T)$, $T$ in Kelvin) for the models BC (top panel), LC
(central panel) and LH (bottom panel) at $t \approx 6.3 \cdot
10^6$y. The black lines are iso-contours of the beam mass fraction
with values $0.1$ (outermost) and $0.9$ (innermost), that correspond
to the limits of the cocoon and the beam, respectively. Please, note
that the colored version of the figure is only provided in the
electronic version of the journal.}
\label{fig:T}
\end{figure*}

\begin{figure*}
\epsfxsize=14cm 
%
\caption{Snapshots of the Lorentz factor $W$ for the models BC (left
panel), LC (central panel) and LH (right panel) at $t \approx 6.6
\cdot 10^6$y for model LH and $t \approx 6.3 \cdot 10^6$y for models
LC and BC.  The black lines are iso-contours of the beam mass fraction
with values $0.1$ (outermost) and $0.9$ (innermost), that correspond
to the limits of the cocoon and the beam, respectively. Please, note
that the colored version of the figure is only provided in the
electronic version of the journal.}
\label{fig:W}
\end{figure*}

\begin{figure*}
%
\caption{Profiles along the jet axis ($r=0$) at $t \approx 6.3 \cdot
10^6$y of (a) density (in units of $\rho_m$), (b) pressure (in units
of $\rho_mc^2$), (c) specific internal energy density (in units of
$c^2$), (d) local sound speed (in units of $c$), (e) temperature (in
Kelvins), (f) Lorentz factor and (g) beam mass fraction. The units
given at the abscissas are in kpc.}
\label{fig:profiles}
\end{figure*}

\begin{figure*}
%
\caption{Time evolution of (a) the jet length $l_j$, average cavity radius
$r_c$, average cocoon radius $r_l$, (b) the aspect ratios
(length/radius) of cocoon and cavity, (c) the hotspot pressure
$P_{hs}$, average cavity pressure $P_c$ and (d) jet velocity $v_j$ for
models BC (dashed), LC (solid) and LH (dotted), respectively.  The jet
velocities of models LC and LH have been divided by 10 and 100,
respectively, in order to enhance the readability of the figure.}
\label{fig:logevolution}
\end{figure*}

\begin{figure*}
%
\caption{Bow shock velocity $v_c$ for the three models. }
\label{fig:vevolution}
\end{figure*}
\begin{figure*}
%
\caption{Hotspot pressure $P_{hs}$ (solid) and ram pressure
$\rho_m v_c^2$ of the bow shock (dashed).} \label{fig:pevolution}
\end{figure*}

\begin{figure*}
%
\caption{Ratio of beam particles to external particles in the cocoon.}
\label{fig:nratio}
\end{figure*}

\begin{figure*}
\epsfxsize=17.5cm 
%
\caption{Simulated total intensity (in logarithmic arbitrary units)
radio maps for models BC, LC and BH (from top to bottom) after $6
\cdot 10^6\,$y, and for a spectral index of $-0.7$. Both, the jet (to
the right) and counter-jet emission are shown for a viewing angle of
$45^\circ$ in all three models. To allow for a better comparison with
observed sources the images have been convolved with a circular
Gaussian beam with FWHM of 10$R_b$. The same limiting ``noise'' level 
is used in all frames to allow for a better comparison of the models.}  
\label{fig:radiomap}
\end{figure*}

\begin{figure*}
%
\caption{Surface brightness per unit of frequency for model BC at
$10\,$keV (top) and $10\,$MeV (bottom), respectively. No plots are
shown for the other two models, as the plots for model LH and the plot
of model LC at $10\,$keV would be indistinguishable from those of
model BC, and as model LC shows no bremsstrahlung emission.  at
$10\,$MeV. The colour scales are normalised to maximum values of $4.5
\cdot 10^{-25}\,$erg$\,$sec$^{-1}$Hz$^{-1}\,$cm$^{-2}$ (top) and $2.8
\cdot 10^{-29}\,$erg$\,$sec$^{-1}$Hz$^{-1}\,$cm$^{-2}$ (bottom),
respectively. Please, note that the colored version of the figure is
only provided in the electronic version of the journal.}
\label{fig:bremss_90_bk}
\end{figure*}

\begin{figure*}
%
\caption{Beam velocity averaged over cross sectional cuts through the
beam perpendicular to the jet axis for different grid resolutions. The
solid line corresponds to a resolution of 6 zones per beam radius
(working resolution). The dashed and dashed--dotted lines correspond
to test resolutions of 4 and 10 zones per beam radius, respectively.}
\label{fig:resdep}
\end{figure*}

\end{document}